\documentclass[preprint,authoryear,12pt]{elsarticle}

\usepackage{color}
\usepackage{bm}
\usepackage{amsfonts}
\usepackage{algorithmic}
\usepackage{algorithm}
\usepackage{booktabs}
\usepackage{multirow}

\usepackage{graphicx}
\usepackage{subfigure}
\usepackage{epsfig}

\usepackage{amssymb}

\newcommand{\tab}[4]
{
\begin{table}[htbp]
  \centering
  \caption{#2}
  \label{#1}
  \begin{tabular}{#3}
  #4
  \end{tabular}
\end{table}
}

\newcommand{\tf}[1]
{
\textbf{#1}
}

\newcommand{\fig}[4]
{
\begin{figure}[!htp]
\centering
\includegraphics[angle=0, width=#3\textwidth]{#4}
\caption{#2}\label{#1}
\end{figure}
}

\journal{Journal of Food Engineering}

\begin{document}

\begin{frontmatter}


\title{Qualitative detection of oil adulteration with machine learning approaches}


\author[a]{Xiao-Bo Jin\corref{cor1}}
\ead{jxb9801@163.com}
\author[b]{Qiang Lu}
 \ead{qlu@swu.edu.cn}
\author[a]{Feng Wang}
\ead{wangfeng$\_$scu@163.com}
\author[c]{Quan-gong Huo}
\ead{qghuo@haut.edu.cn}
\address[a]{Key Laboratory of Processing and Control on Grain Information, School of Information Science and Engineering, Henan University of Technology, Zhengzhou 450001, China}
\address[b]{Citrus Research Institute, SouthWest University, Chongqing 400712, China}
\address[c]{School of Cereals, Oils and Foodstuffs, Henan University of Technology, Zhengzhou 450001, China}
\cortext[cor1]{Corresponding author. Tel.: +086-0371-67756527 Postal Address: Key Laboratory of Processing and Control on Grain Information (Ministry of Education), School of Information Science and Engineering, Henan University of Technology, Lianhua Street, Gaoxin District, Zhengzhou, China (450001)}

\begin{abstract}
The study focused on the machine learning analysis approaches to identify the adulteration of 9 kinds of edible oil qualitatively and answered the following three questions: Is the oil sample adulterant? How does it constitute? What is the main ingredient of the adulteration oil? After extracting the high-performance liquid chromatography (HPLC) data on triglyceride from 370 oil samples, we applied the adaptive boosting with multi-class Hamming loss (AdaBoost.MH) to distinguish the oil adulteration in contrast with the support vector machine (SVM). Further, we regarded the adulterant oil and the pure oil samples as ones with multiple labels and with only one label, respectively. Then multi-label AdaBoost.MH and multi-label learning vector quantization (ML-LVQ) model were built to determine the ingredients and their relative ratio in the adulteration oil. The experimental results on six measures show that ML-LVQ achieves better performance than multi-label AdaBoost.MH.
\end{abstract}

\end{frontmatter}



\section{Introduction}
Oil pureness is a very important aspect of the quality edible oil for special reasons of the sensory properties, perceived health values and the confidence of the health foods \cite{Lees2003}. Adulteration of oil products, involving the replacement of expensive ingredients with the cheaper substitutes, could potentially be very lucrative for a vendor or raw material supplier. Thus, continuous vigilance is required to control the adulteration of the edible oil products such as sesame oil in Asian food and to protect the interests of the consumers.

In order to evaluate the quality of the edible oil and detect its adulteration, a number of chromatographic \cite{Isabel2002,Carlo2006} and spectroscopic methods, including fluorescence \cite{Ewa2005}, near-infrared (NIR) \cite{Alfred2004,Gerard2002}, Fourier transform infrared spectroscopy (FT-IR) \cite{Ruben2010,Abdelkhalek2012}, FT-Raman \cite{Hong2005}, nuclear magnetic resonance (NMR) \cite{Vigli2003,Georgia2005,Alexia2010}, mass spectrometry (MR) \cite{Isabel2002,Maria2008}, mid-infrared spectroscopy (MIR) \cite{Gozde2009,Abdul2011}, dielectric spectroscopy \cite{Hu2010} and high-performance liquid chromatography (HPLC) \cite{Dionisi1995,Ali1995,Cosima2010,Luiz2012} are widely used to analyze the composition of the oil (e.g. olive oil or sun-flower oil) even the possible adulterants.  In most applications found in the literature of oil-adulteration detection, the multivariate statistical analysis like linear discriminant analysis (LDA) \cite{Isabel2002,Abdul2011}, principal component analysis (PCA) \cite{Alfred2004}, partial least square regression (PLS) \cite{Cataldo2012,Gerard2002,Gozde2009,Abdul2011} and artificial neural networks (ANN) \cite{Concepcion2002} are applied to further analyze the oil spectroscopy.

However, these data analysis techniques heavily depend on the chromatographic and spectroscopic methods or the hand-crafting analysis methods. In the case of spectral data, not all contributes are unique or useful information although the elimination of predictors of limited or negligible utility can increase the efficiency of the models or make their interpretation simpler. Some attribute features only can be suitable for certain chromatographic or spectroscopic methods but not for others. Generally, the conventional analysis approaches depend on the visualization analysis of the special features and the special chemical treatment for certain oil samples. It is worthy to develop the powerful data analysis approaches under more precise measures for the authentication of the adulterant oil.

In this study, we focus on detecting the adulteration qualitatively and precisely with machine learning methods including classification method and multi-label learning. After extracting HPLC data based on the triglyceride of oil samples, we tried to solve the following three questions simultaneously: Is the oil sample adulterant? How does it constitute? What is the main ingredient of the adulteration oil? First, we used the classical adaptive boosting with multi-class Hamming loss (AdaBoost.MH) classification method to determine that the oil is adulterant or not in contrast with the support vector machine (SVM) that is widely applied in the oil authentication \cite{Sonia2007,Olivier2009}. AdaBoost.MH is the combination of multiple weak classifiers, where each weak classifier related to an optimal feature implements the feature selection in some sense. The experimental results show that AdaBoost.MH achieves the superior performance than SVM. Meanwhile, it is easy-to-use for the practical application with only one parameter $T$ and $T$ is set to $100$ enough to work well unlike SVM that implements the model optimization in the grid of two parameters. In addition, we also explored the role of the principal component analysis (PCA) preprocessing in AdaBoost.MH classification method and the empirical results demonstrate that AdaBoost.MH performs better when using all features because PCA doesn't consider the class information and can not guarantee the class separation. Second, we regarded the samples of the pure oil and the adulterant oil as ones with one label and multiple labels, respectively. Then we could apply multi-label learning to detect the ingredients of the adulterant oil. We further analyzed the relative ratio of the components in the adulterant oil with multi-label learning vector quantization (ML-LVQ) \cite{Xiaobo2012} although the multi-label AdaBoost.MH can predict the labels of the oil samples \cite{Quangong20122}. Finally, The statistical analysis including the micro-F1, the macro-F1, the average precision, the one-error, the accuracy and the detect rate measures gives the full comparisons between the multi-label AdaBoost.MH and the ML-LVQ approaches.

\section{Materials and methods}
\subsection{Reagent and Chemicals}
The acetonitrile and dichloromethane were obtained from Merck (Darmstadt, Germany) for the gradient elution. Water was purified on a Milli-Q system (Millipore, Bedford, MA, USA). All other reagents were of analytical grade unless otherwise stated.

\subsection{Sample preparation}
Nine kinds of edible vegetable oil was collected from Henan province and Shandong province (China), including the soybean oil, the palm oil, the sesame oil, the corn oil, the peanut oil, the sunflower oil, the rice bran oil, the rap oil and the cotton oil (see Tab. 1). The adulterant oil is composed of 4 kinds of basis oil (soybean, peanut, sunflower and sesame), which is adulterated with other 5 kinds of oil in the specified percentage range between $5\%$ and $99\%$. The dataset contains 370 examples with 1607 dimensions associated with a series of time stamp.

\tab{tab:distribution}{Category distribution of the edible oil: \& denotes the mixture of both}{llc}{ \hline
No. & Edible oil & No. of examples \\
\hline
0 & soybean & 34 \\
1 & peanut & 39 \\
2 & sunflower & 17 \\
3 & corn  & 10 \\
4 & palm & 27 \\
5 & sesame & 37 \\
6 & cotton & 0 \\
7 & rap & 58 \\
8 & rice bran & 24\\
 \hline
9& soybean\&sesame & 21\\
10 & soybean\&palm & 9 \\
11& soybean\&corn & 3 \\
12& soybean\&sunflower & 3 \\
13& soybean\&peanut & 9 \\
14& sunflower\&sesame & 21 \\
15& palm\&sesame & 9 \\
16& peanut\&sesame & 20\\
17& peanut\&palm & 9\\
18& peanut\&corn & 2\\
19& peanut\&sunflower & 9\\
20& sesame\&cotton & 9\\
\hline
 }

\subsection{Instrumentation and chromatographic conditions}
The chromatogram data of edible oil was obtained by the HPLC system including a vacuum degasser, an auto-sampler and a binary pump from Agilent Series 1100 (Agilent Technologies, Santa Clara, CA). The HPLC system was equipped with a reversed phase $C_{18}$ analytical column of 250 mm $\times$ 4.6 mm (5 $\mu m$ particle size). Column temperature was kept at 70$^{\circ}$C. The injected sample volume was 20$\mu$L. Mobile phases A and B was acetonitrile and dichloromethane (35:65), respectively. The flow rate was set to 1.00$\mu$L/min.

\subsection{Data analysis}
In our work, we apply three sorts of methods to analyze the oil samples in the framework of machine learning. First, we apply the classification method to discriminate the adulterant oil from the pure oil. Further, we regard the pure oil as the sample with only one label and the adulterant oil as one with multiple labels and then resort to the multi-label learning method. It can not only judge whether the oil is pure or adulterant but also recognize the composite ingredients of the adulterant oil effectively. Finally, we use ML-LVQ to analyze onward the relative ratio of the different component in the adulterant oil.

As a reference, we also give the analysis results on the state-of-art methods including SVM and PCA, where SVM \cite{Burges1998} constructs a hyperplane or set of hyperplanes in the high dimensional space to maximize the margin. We built PCA extraction, SVM classifier and AdaBoost.MH classifier with Matlab language on Matlab 2011b platform (Mathworks Inc. U.S.), and multi-label AdaBoost.MH\& ML-LVQ with Java language on JDK 1.6 (Oracle, Inc. U.S.).

\subsubsection{Adulteration detection with binary AdaBoost.MH}\label{sec:binary_ada}
The classic AdaBoost.MH \cite{Schapire1999} is introduced as follows. Let us consider a labeled dataset $D = \{(\bm{x}_{n},t_{n})\}_{n=1}^{N}$, where $\bm{x}_{n} \in R^{d}$ and $t_{n} \in \{+1,-1\}$. We often use $+1$ and $-1$ to sign the oil sample as the adulterant oil and the pure oil, respectively. The following classification model is built by using a non-linear function from the input data to the real number. By the sign of $f(\bm{x})$, we can predict the unknown sample $\bm{x}$ as the adulteration if $f(\bm{x}) > 0$ otherwise as the pure oil.

AdaBoost.MH \footnote{Schapire \cite{Schapire1999} proposed the discrete AdaBoost.MH and the real AdaBoost.MH in his work, but we really refer to the real AdaBoost.MH in the context. } combines the final hypothesis by calling the weak learner repeatedly in a series of rounds in a varying data distribution. On each round, the weights of each incorrectly classified example are increased, and the weights of each correctly classified example are decreased, so next new classifier focuses on the examples which have so far eluded correct classification.

\subsubsection{Detection of ingredients in non-pure oil with multi-label AdaBoost.MH}
In the framework of multi-label classification, we can regard the oil sample as the example with multiple labels. So the pure oil and the adulterant oil are attached with only one label and more than one label, respectively. The multi-label classification algorithm can not only recognize the components of the adulterant oil, but also detect the adulteration oil like the binary AdaBoost.MH in section \ref{sec:binary_ada} when predicting the unknown oil sample. We will introduce multi-label AdaBoost.MH algorithm \cite{Schapire1999} in the following.

Let us consider a labeled dataset $D = \{(\bm{x}_{n},t_{n})\}_{n = 1}^{N}$, where $t_{n}$ is the subset of $\mathcal{L} = \{1,2,\cdots,L\}$ instead of $\{+1,-1\}$. The successful multi-label learning will produce a ranking of the possible labels to rank the outputs in the label set $t$ on the top of those not in $t$ or make the difference between the predict label set and the true label set as small as possible.

Multi-label AdaBoost.MH algorithm decomposes the problem into $k$ binary classification problems. We can take the predicting target as $L$ binary labels depending on whether a label $l$ is or not included in $t_{n}$. Then $f(\bm{x},l)$ can be viewed as one of $L$ binary predictions for the label $l$. On round $t$, the weak learner accepts a weighted dataset with a distribution and generates a weak classifier. We will assign the instance $\bm{x}$ with the label $l$ only if $f(\bm{x},l) > 0$.

\subsubsection{Analysis of the relative ratio on components with ML-LVQ}
It is worthy to note that multi-label AdaBoost.MH can predict the constituents of the adulterant oil but it can not determine the relative ratio of the ingredient effectively in the adulterant oil. For instance, although we know that the adulterated oil is composed of the soybean oil and peanut oil, it is not clear that the soybean oil is adulterated in the peanut oil or the peanut oil is adulterated in the soybean oil. ML-LVQ \cite{Xiaobo2012} considers the label ranking problem effectively when the example is assigned with multiple labels.

ML-LVQ extends LVQ to handle the multi-label problems, where the minimum classification error on the set of labels can approximate the rank loss. ML-LVQ minimizes the upper bound of ranking loss by the stochastic gradient descent. When the n-th training example is given in the t-th iteration, two prototypes from the positive class and the negative class are updated for each label from the set $t_n$ by computing the derivative of the loss.

Given a test example $\bm{x}$, the algorithm will output the predict value for each label. The meta-labeler with the meta-model predicts the number of $\bm{x}$'s labels as $k(\bm{x})$. So the top $k(\bm{x})$ highest scoring categories are chosen as the labels of $\bm{x}$ for prediction and then sorted to predict their relative ratio in the oil sample.

\section{Results and Discussion}
 For 370 oil examples, the evaluation measures of all algorithms on the oil dataset were obtained via 10 runs of 5-fold cross validation to avoid the randomness introduced by splitting the dataset. The detailed procedure was below:
\begin{enumerate}
  \item
  For each run, the dataset was randomly divided into five disjoint subsets of approximately same size by stratified sampling. We kept the same divisions for all learning algorithms.
  \item
  Each of five subsets was used as test set and the remaining data was used for training. The five subsets were used for testing rotationally for evaluating the performance of machine learning methods.
  \item
  During each training process, the training parameters were determined as follows: first, we held out 1/3 of the training data by stratified sampling for validation while the model parameters were estimated on the remaining 2/3 of data (the split of training data is the same for all learning algorithms). After selecting training parameters that gave the highest validation accuracy, all the training data was used to re-train the classifier for evaluation on test data.
\end{enumerate}

\subsection{Overview of HPLC spectra of edible oil}
Fig. 1 shows the chromatographic data of two kinds of pure oil and their mixture. There is much overlap among these samples except a slight difference in the weave peak. The similarity of the spectra makes the detection with the hand-crafting features so difficult, and then we resort to machine learning techniques to detect the difference.

 \fig{fig:sprectral}{Normalized chromatographic image for peanut oil, soybean oil and their hybrid (with $60\%$ peanut oil)}{0.55}{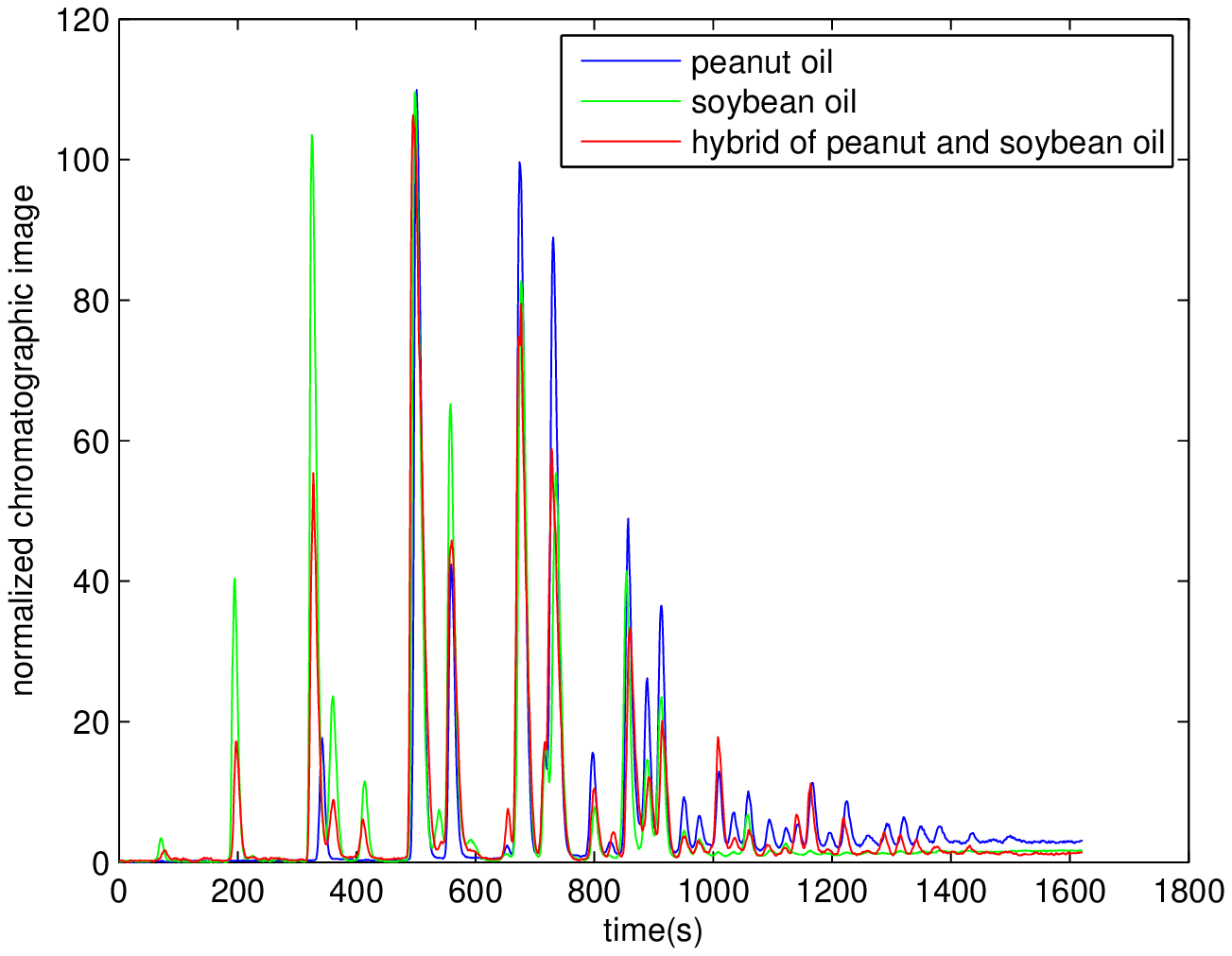}

\subsection{Evaluation Measures}
The accuracy is a popular measure used widely in the classification algorithm to evaluate the performance. As for the evaluation of the multi-label algorithms, we use both of the bipartitions and the rankings \cite{Tsoumakas2010} with respect to the ground truth of multi-label data. The bipartitions measures compute the average difference between the actual and the predicted set of the labels including the macro-averaging and the micro-averaging of F1 (mac-F1 and mic-F1). The ranking measures evaluate the average difference between the true ranking and the predicted ranking including the one-error and the average precision (avg-prec).

We also give the accuracy measure in order to contrast with previous binary classification problems. Finally, we provide the detect rate (detect-rate) to evaluate the predicted label set is identical to the true label set or not.

\subsection{Detection results with AdaBoost.MH classification model}
SVM is a classic classification approaches applied in chemometrics \cite{Cogdill2004} and diagnosis \cite{Comak2007}. In our work, we give the comparisons on SVM and AdaBoost.MH. Specially, we run AdaBoost.MH with and without PCA preprocessing to analyze the effects of PCA.

\fig{fig:svm_para}{Accuracy of SVM with the different model parameters on the validation dataset: the optimal combination of $(\log_2 C, \log_2 \gamma)$ was found at the value $(1,5)$, which is signed with the square point and slightly superior than its rival ones.}{0.55}{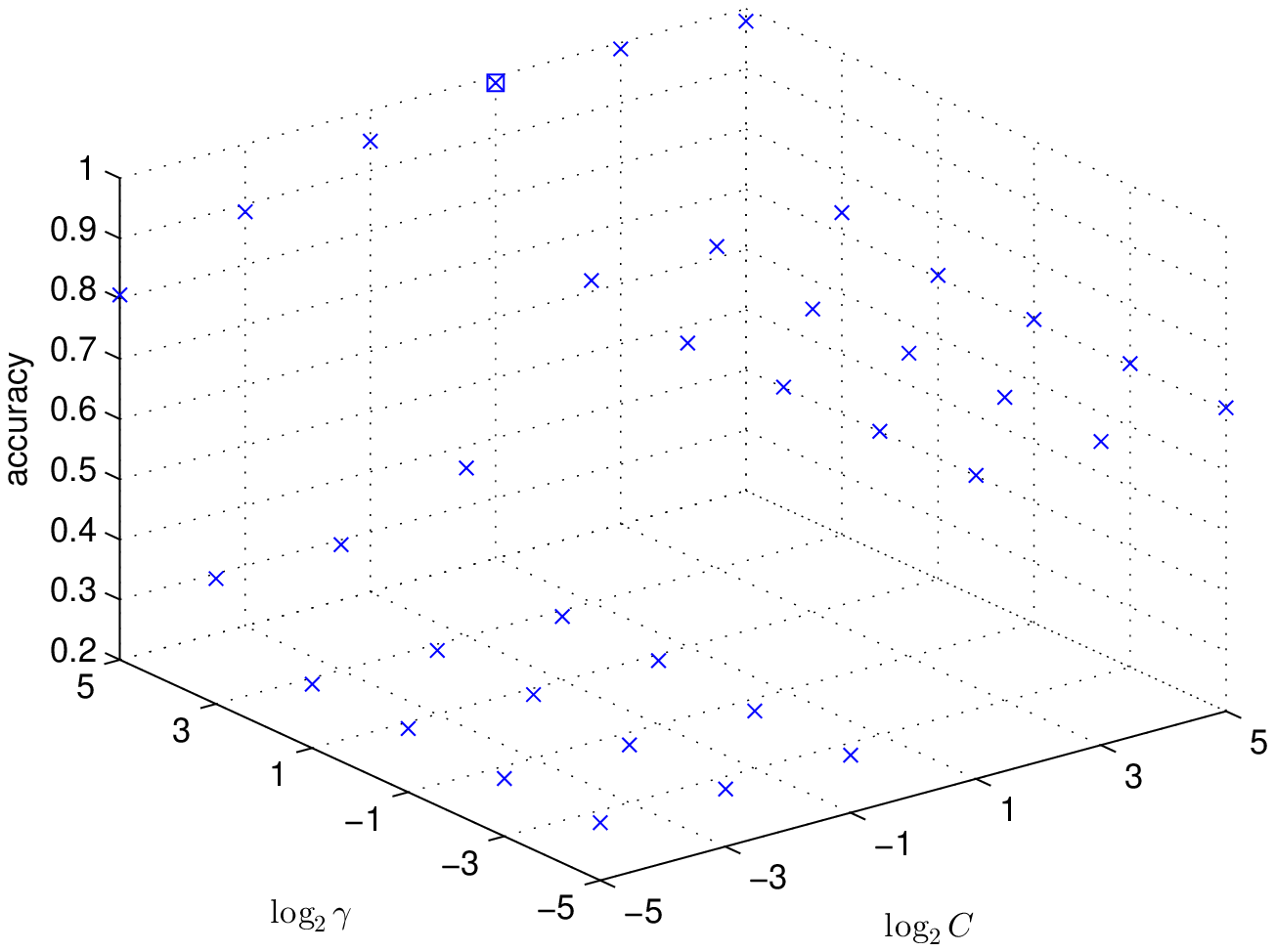}

The optimal model parameters were found with the strategy described in the beginning of this section. We chosen $T$ from $\{100,200,300,400,500\}$ for AdaBoost.MH, where $T$ is the number of stumps. The SVM classifier with RBF kernel was implemented by the bioinformatics toolbox in Matlab. We only considered the tradeoff parameter $C$ and the kernel width $\gamma$, where both of $\log_{2}C$ and $\log_{2}\gamma$ were selected from $\{-5,-3,-1,1,3,5\}$. The best average accuracy with AdaBoost.MH and SVM on ten run of 5-cross fold are $96.46\%$ and $94.92\%$, respectively. We can see that AdaBoost.MH achieves better performance than SVM with the RBF kernel. The optimal model parameters from one of ten runs are listed in Tab. 2. In Fig. 2, we can find that fact that the accuracy is a non-linear function of the parameters ($\log_{2}C$ and $\log_{2}\gamma$) makes the determination of the model parameters difficult. But for AdaBoost.MH (see Fig. 3), the accuracy will generally increase with the growing parameter $T$ until convergence. The ideal model parameter is easily determined when considering the tradeoff between the running time and the accuracy. Tab. 2 shows that AdaBoost.MH algorithm converges early before $T$ reaches its maximum.

\begin{table}
  \centering
  \caption{Optimal parameters are found by SVM and AdaBoost.MH in a run, where SVM and AdaBoost.MH achieves the average accuracy $96.22\%$ and $96.76\%$, respectively.}\label{tab:valid_para}
  \begin{tabular}{lccc}
    \hline
   \multirow{2}{*}{Fold No.} & \multicolumn{2}{c}{SVM} & AdaBoost.MH \\
   \cmidrule(r{0.5em}){2-3}  \cmidrule(l{0.5em}){4-4}
    & $\log_2 C$ & $\log_2 \gamma$  & $T$ \\
    \hline
    1 & 1  & 5  &  200 \\
    2 & 3  & 5  &  100 \\
    3 &  3 & 5  &  300 \\
    4 &  1 & 5 &  100 \\
    5 & 1  & 5 &  200 \\
    \hline
  \end{tabular}
\end{table}

\fig{fig:rmh_para}{Accuracy of AdaBoost.MH varies with $T$ on the validation dataset}{0.55}{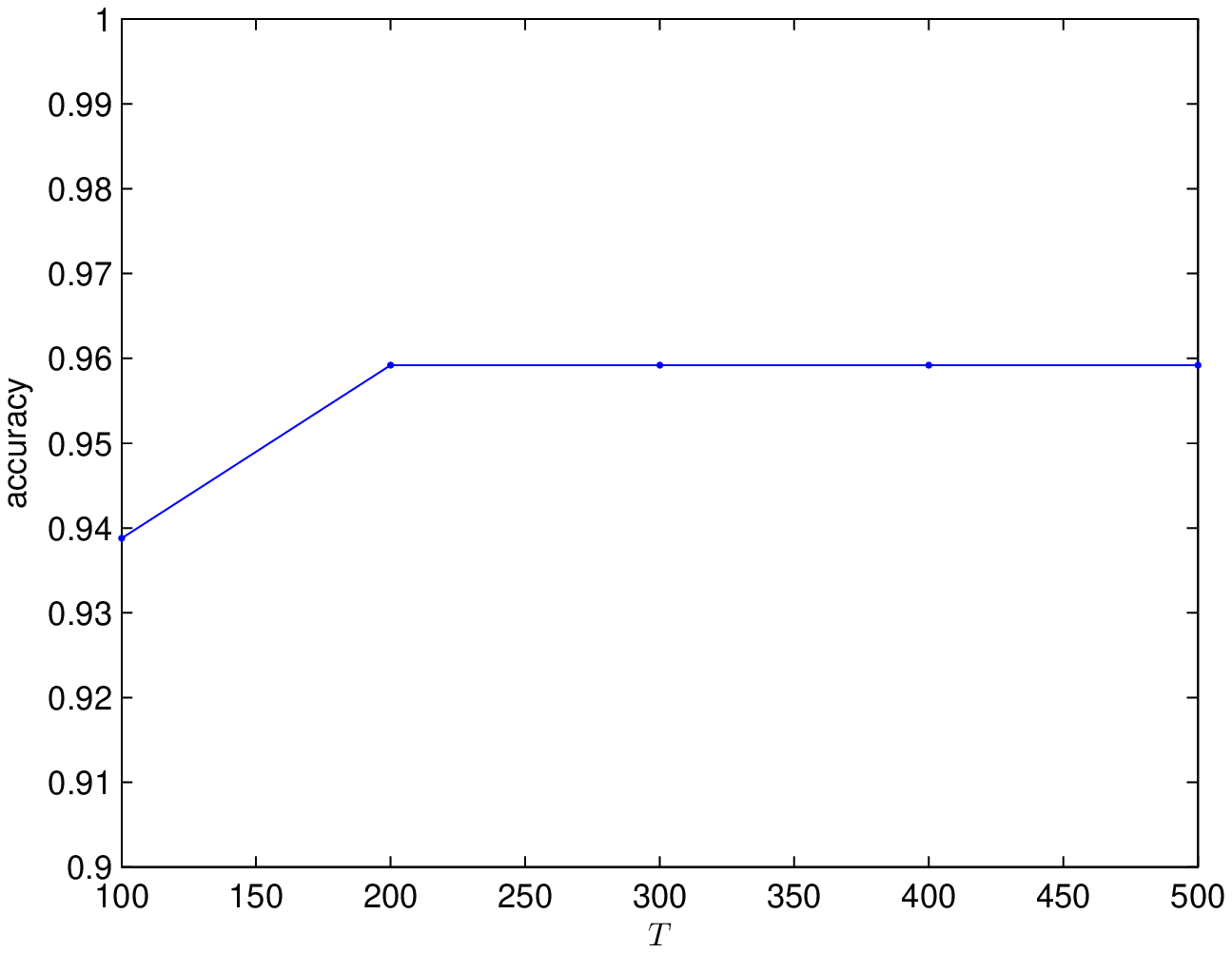}

Finally, we also explored how PCA influences the classifier (see AdaBoost.MH in Tab. 3). We can achieve the high accuracy ($90.05\%$) with 35 dimensions of features, but it is far inferior to the performance using all features due to the facts that PCA can not guarantee the class separation on the features whiling preserving most of the data information. In reality, AdaBoost.MH implements the feature selection in some sense.

\tab{tab:pca_results}{AdaBoost.MH results with PCA preprocessing under the different cumulative variance (the last column but one shows the result when selecting all components such that the eigenvalue is large than 0.)}{lccccc}{
\hline
Variance & $95\%$ & $98\%$ & $99\%$ & $>0$ & all \\
\hline
Dimensions & $15$ & $25$ & $35$ &  $948$ & $1607$ \\
Results & $88.41\%$ & $89.38\%$ & $90.05\%$ & $84.76\%$ & $96.46\%$ \\
\hline
}

\subsection{Detection results with multi-label learning algorithm}
We compared the performance of multi-label AdaBoost.MH and ML-LVQ algorithm on both of the detection of the ingredient and the relative ratio of the adulterant oil in Tab. 4. We used the macro-F1, the micro-F1, the one-error, the avg-prec, the accuracy and the detect-rate for the performance evaluation. In the validation of the model parameters, we observed that it is no obvious difference when using other measures instead of mic-F1 for the validation.

In the implementation of ML-LVQ, the prototypes were initialized by K-means clustering of class-wise data. Each attribute of the examples was scaled to $[-1,+1]$. We only optimized the number of the prototypes (S) for ML-LVQ and the number of the stumps (T) for multi-label AdaBoost.MH, where $S$ was set to $\{1,3,5,7,9\}$ and $T$ to $\{20,40,60,80,100\}$. The initial learning rate $\eta(0)$ in the stochastic gradient descent was assigned to 0.1*cov, where cov is the average distance of all training examples to the nearest cluster center. By default, $M = 40$ and $\alpha = 0$ is enough to work well. ML-LVQ used AdaBoost.MH classifier with 100 decision stumps as the meta-labeler.

\tab{tab:multi_label}{Performance of multi-label AdaBoost.MH and ML-LVQ on the oil samples ($\%$):the best measure is highlighted in boldface}{lcc}{
\hline
Measurers & AdaBoost.MH & ML-LVQ \\
 \hline
 Mac-F1 & 91.18 & \tf{94.46}\\
 Mic-F1 & 95.39 & \tf{97.09}\\
 One-error & \tf{1.14} & \tf{1.14}\\
 Avg-prec & 98.48 & \tf{98.55}\\
 Accuracy & 88.08 & \tf{95.32}  \\
 Detect-rate & 88.76 &  \tf{92.59} \\
 \hline
}

In Tab. 4, we observe that ML-LVQ (accuracy = 95.32\%) obtains higher accuracy than multi-label AdaBoost.MH (accuracy = $88.08\%$) and it is slightly inferior to AdaBoost.MH classification method (with $96.46\%$). Further, ML-LVQ shows the superior performance in predicting the ingredient components and the relative ratio of components in contrast with multi-label AdaBoost.MH especially on the mac-F1 measure. It is surprising that the predicting ability of ML-LVQ can also reach to $92.59\%$ (refer to the detect-rate) when the algorithm predicts the entire set of the ingredients. Finally, Tab. 5 gives the model parameters found by multi-label AdaBoost.MH and ML-LVQ in one run, where ML-LVQ always sets $S = 1$ as the optimal parameter. We need few prototypes enough to represent a kinds of examples since most kinds of oil examples contains few examples such that nine cotton oil samples only appear in the mixture of the sesame oil and the cotton oil.

\tab{tab:multi-label-valid}{Model selection in validation dataset on multi-label AdaBoost.MH (with micro-F1 = $95.24\%$) and ML-LVQ algorithm (with micro-F1 = $97.40\%$) in a running}{lcc}{
\hline
Fold No. & $T$ (AdaBoost.MH) & $S$ (ML-LVQ) \\
\hline
1 & 40 & 1\\
2 & 80 & 1\\
3 & 80 & 1\\
4 & 60 & 1\\
5 & 60 & 1\\
\hline
}

Let us take a close look at 25 incorrectly predicted examples by ML-LVQ from 370 examples. Fig. 4 gives the detect rates when adulterating the oil with the ratio of both ingredients in the range $[0\%,99\%]$. It is clear to see that the detect rates will decrease when increasing the relative ratio of both ingredients. Then the multi-label algorithm can predict more precisely for nearly $50\%$ than nearly $0\%$ or $100\%$ (the point $0\%$ corresponds to the case of the pure oil). In fact, our multi-label algorithm achieves about $98\%$ in the detect-rate measure for 246 pure oil samples in Fig. 4. There appears the vibration in the point $95\%$ (the square point in Fig. 4) since there are 3 examples for $95\%$ adulteration rate without demonstrating the statistical meanings. Fig. 4 shows the phenomenon that is consistent with our intuition.

\fig{fig:error}{The detect-rate varies with the adulteration ratio in a 5-cross-fold validation (detect-rate = $93.24\%$).}{0.55}{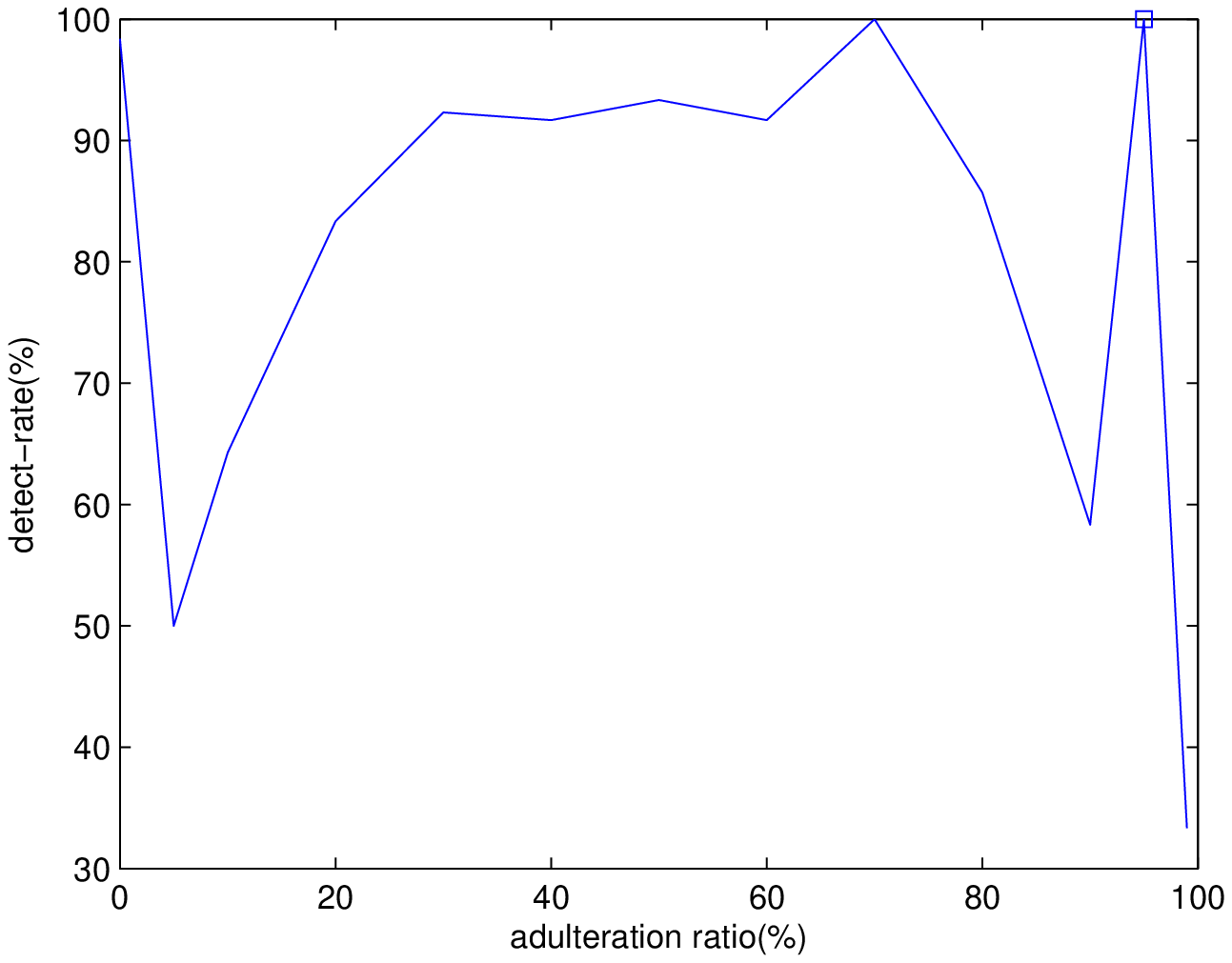}

Finally, we compare the true labels and the predict labels of 25 mis-classified examples to discover that the ML-LVQ can predict the main ingredient for all of them. For instance, the algorithm predicts the certain sample incorrectly as the palm oil adulterated with the sesame one but the sample is really composes of the palm oil adulterated with $20\%$ soybean one.

\section{Conclusions}
HPLC data of the edible oil with their mixture was investigated to answer the following questions:  Is the oil sample adulterant? How does it constitute? What is the main ingredient of the adulteration oil? In our study, we applied the handy AdaBoost.MH to detect the oil sample is adulterant or not to achieve a higher accuracy than the state of art SVM. In addition, AdaBoost.MH with PCA does not show the better performance than one without any preprocessing since PCA can not guarantee the class separation on the features. Finally, we used multi-label AdaBoost.MH and ML-LVQ to recognize the ingredient and their relative ratio when taking the adulterant oil and the pure oil as the examples with multiple labels and one labels, respectively. The comparisons between multi-label AdaBoost.MH and ML-LVQ on multiple measures show that ML-LVQ is more promising than AdaBoost.MH in solving above three questions. We also note that our method can only detect the oil adulteration qualitatively unlike the least squares support vector machine (LS-SVM) \cite{Wu2008} which can measure the ingredients quantitatively. Finally, these data analysis approaches are also suitable for following other chromatographic and spectroscopic methods such as NIR or MR instead of HPLC.

\section{Acknowledgements}
This work is supported in part by the National Natural Science Foundation of China (NSFC) under Grant No. 61103138, No. U1204617 and the Innovative Funding of Henan University of Technology.


\begin{thebibliography}{32}
\expandafter\ifx\csname natexlab\endcsname\relax\def\natexlab#1{#1}\fi
\providecommand{\bibinfo}[2]{#2}
\ifx\xfnm\relax \def\xfnm[#1]{\unskip,\space#1}\fi
\bibitem[{Agiomyrgianakia et~al.(2010)Agiomyrgianakia, Petrakisb \&
  Daisa}]{Alexia2010}
\bibinfo{author}{Agiomyrgianakia, A.}, \bibinfo{author}{Petrakisb, P.~V.}, \&
  \bibinfo{author}{Daisa, P.} (\bibinfo{year}{2010}).
\newblock \bibinfo{title}{Detection of refined olive oil adulteration with
  refined hazelnut oil by employing nmr spectroscopy and multivariate
  statistical analysis}.
\newblock {\it \bibinfo{journal}{Talanta}\/},  {\it \bibinfo{volume}{80}\/},
  \bibinfo{pages}{2165--2171}.
\bibitem[{Brandao et~al.(2012)Brandao, Braga \& Suarez}]{Luiz2012}
\bibinfo{author}{Brandao, L. F.~P.}, \bibinfo{author}{Braga, J. W.~B.}, \&
  \bibinfo{author}{Suarez, P. A.~Z.} (\bibinfo{year}{2012}).
\newblock \bibinfo{title}{Determination of vegetable oils and fats adulterants
  in diesel oil by high performance liquid chromatography and multivariate
  methods}.
\newblock {\it \bibinfo{journal}{Journal of Chromatography A}\/},  {\it
  \bibinfo{volume}{1225}\/}, \bibinfo{pages}{150--157}.
\bibitem[{Burges(1998)}]{Burges1998}
\bibinfo{author}{Burges, C. J.~C.} (\bibinfo{year}{1998}).
\newblock \bibinfo{title}{A tutorial on support vector machines for pattern
  recognition}.
\newblock {\it \bibinfo{journal}{Journal Data Mining and Knowledge
  Discovery}\/},  {\it \bibinfo{volume}{2}\/}, \bibinfo{pages}{121--167}.
\bibitem[{Caetano et~al.(2007)Caetano, \"{U}st\"{u}n, Hennessy,
  Smeyers-Verbeke, Melssen, Downey, Buydens \& Heyden}]{Sonia2007}
\bibinfo{author}{Caetano, S.}, \bibinfo{author}{\"{U}st\"{u}n, B.},
  \bibinfo{author}{Hennessy, S.}, \bibinfo{author}{Smeyers-Verbeke, J.},
  \bibinfo{author}{Melssen, W.}, \bibinfo{author}{Downey, G.},
  \bibinfo{author}{Buydens, L.}, \& \bibinfo{author}{Heyden, Y.~V.}
  (\bibinfo{year}{2007}).
\newblock \bibinfo{title}{Geographical classification of olive oils by the
  application of cart and svm to their ft-ir}.
\newblock {\it \bibinfo{journal}{Journal of . Chemometrics}\/},  {\it
  \bibinfo{volume}{21}\/}, \bibinfo{pages}{324--334}.
\bibitem[{Calvano et~al.(2010)Calvano, Aresta \& Zambonin}]{Cosima2010}
\bibinfo{author}{Calvano, C.~D.}, \bibinfo{author}{Aresta, A.}, \&
  \bibinfo{author}{Zambonin, C.~G.} (\bibinfo{year}{2010}).
\newblock \bibinfo{title}{Detection of hazelnut oil in extra-virgin olive oil
  by analysis of polar components by micro-solid phase extraction based on
  hydrophilic liquid chromatography and maldi-tof mass spectrometry}.
\newblock {\it \bibinfo{journal}{Journal of Mass Spectrometry}\/},  {\it
  \bibinfo{volume}{45}\/}, \bibinfo{pages}{981--988}.
\bibitem[{Cataldo et~al.(2012)Cataldo, Piuzzi, Cannazza \&
  Benedetto}]{Cataldo2012}
\bibinfo{author}{Cataldo, A.}, \bibinfo{author}{Piuzzi, E.},
  \bibinfo{author}{Cannazza, G.}, \& \bibinfo{author}{Benedetto, E.~D.}
  (\bibinfo{year}{2012}).
\newblock \bibinfo{title}{Classification and adulteration control of vegetable
  oils based on microwave reflectometry analysis}.
\newblock {\it \bibinfo{journal}{Journal of Food Engineering}\/},  {\it
  \bibinfo{volume}{112}\/}, \bibinfo{pages}{338 -- 345}.
\bibitem[{\c{C}omaka et~al.(2007)\c{C}omaka, Arslana \& \.{I}brahim
  T\"{u}rko\u{g}lub}]{Comak2007}
\bibinfo{author}{\c{C}omaka, E.}, \bibinfo{author}{Arslana, A.}, \&
  \bibinfo{author}{\.{I}brahim T\"{u}rko\u{g}lub} (\bibinfo{year}{2007}).
\newblock \bibinfo{title}{A decision support system based on support vector
  machines for diagnosis of the heart valve diseases}.
\newblock {\it \bibinfo{journal}{Computers in Biology and Medicine}\/},  {\it
  \bibinfo{volume}{37}\/}, \bibinfo{pages}{21--27}.
\bibitem[{Christy et~al.(2004)Christy, Kasemsumran, Du \& Ozaki}]{Alfred2004}
\bibinfo{author}{Christy, A.~A.}, \bibinfo{author}{Kasemsumran, S.},
  \bibinfo{author}{Du, Y.}, \& \bibinfo{author}{Ozaki, Y.}
  (\bibinfo{year}{2004}).
\newblock \bibinfo{title}{The detection and quantification of adulteration in
  olive oil by near-infrared spectroscopy and chemometrics}.
\newblock {\it \bibinfo{journal}{Analytical Sciences}\/},  {\it
  \bibinfo{volume}{20}\/}, \bibinfo{pages}{935--940}.
\bibitem[{Cogdill \& Dardenne(2004)}]{Cogdill2004}
\bibinfo{author}{Cogdill, R.~P.}, \& \bibinfo{author}{Dardenne, P.}
  (\bibinfo{year}{2004}).
\newblock \bibinfo{title}{Least-squares support vector machines for
  chemometrics: An introduction and evaluation}.
\newblock {\it \bibinfo{journal}{Journal of Near Infrared Spectroscopy}\/},
  {\it \bibinfo{volume}{2}\/}, \bibinfo{pages}{93--100}.
\bibitem[{Devos et~al.(2009)Devos, Ruckebusch, Durand, Duponchel \&
  Huvenne}]{Olivier2009}
\bibinfo{author}{Devos, O.}, \bibinfo{author}{Ruckebusch, C.},
  \bibinfo{author}{Durand, A.}, \bibinfo{author}{Duponchel, L.}, \&
  \bibinfo{author}{Huvenne, J.-P.} (\bibinfo{year}{2009}).
\newblock \bibinfo{title}{Support vector machines (svm) in near infrared (nir)
  spectroscopy: Focus on parameters optimization and model interpretation}.
\newblock {\it \bibinfo{journal}{Chemometrics and Intelligent Laboratory
  Systems}\/},  {\it \bibinfo{volume}{96}\/}, \bibinfo{pages}{27--33}.
\bibitem[{Dionisi et~al.(1995)Dionisi, Prodolliet \& Tagliaferri}]{Dionisi1995}
\bibinfo{author}{Dionisi, F.}, \bibinfo{author}{Prodolliet, J.}, \&
  \bibinfo{author}{Tagliaferri, E.} (\bibinfo{year}{1995}).
\newblock \bibinfo{title}{Assessment of olive oil adulteration by
  reversed-phase high-performance liquid chromatography/amperometric detection
  of tocopherols and tocotrienols}.
\newblock {\it \bibinfo{journal}{Journal of the American Oil Chemists¡¯
  Society}\/},  {\it \bibinfo{volume}{72}\/}, \bibinfo{pages}{1505--1511}.
\bibitem[{Downey et~al.(2002)Downey, McIntyre \& Davies}]{Gerard2002}
\bibinfo{author}{Downey, G.}, \bibinfo{author}{McIntyre, P.}, \&
  \bibinfo{author}{Davies, A.~N.} (\bibinfo{year}{2002}).
\newblock \bibinfo{title}{Detecting and quantifying sunflower oil adulteration
  in extra virgin olive oils from the eastern mediterranean by visible and
  near-infrared spectroscopy}.
\newblock {\it \bibinfo{journal}{Journal of agricultural and food
  chemistry}\/},  {\it \bibinfo{volume}{50}\/}, \bibinfo{pages}{5520--5525}.
\bibitem[{El-Hamd \& El-Fizga(1995)}]{Ali1995}
\bibinfo{author}{El-Hamd, A.~H.}, \& \bibinfo{author}{El-Fizga, N.~K.}
  (\bibinfo{year}{1995}).
\newblock \bibinfo{title}{Detection of olive oil adulteration by measuring its
  authenticity factor using reversed-phase high-performance liquid
  chromatography}.
\newblock {\it \bibinfo{journal}{Journal of Chromatography A}\/},  {\it
  \bibinfo{volume}{708}\/}, \bibinfo{pages}{351--355}.
\bibitem[{Fragaki et~al.(2005)Fragaki, Spyros, Siragakis, Salivaras \&
  Dais}]{Georgia2005}
\bibinfo{author}{Fragaki, G.}, \bibinfo{author}{Spyros, A.},
  \bibinfo{author}{Siragakis, G.}, \bibinfo{author}{Salivaras, E.}, \&
  \bibinfo{author}{Dais, P.} (\bibinfo{year}{2005}).
\newblock \bibinfo{title}{Detection of extra virgin olive oil adulteration with
  lampante olive oil and refined olive oil using nuclear magnetic resonance
  spectroscopy and multivariate statistical analysis}.
\newblock {\it \bibinfo{journal}{Journal of agricultural and food
  chemistry}\/},  {\it \bibinfo{volume}{53}\/}, \bibinfo{pages}{2810--2816}.
\bibitem[{Gurdeniz \& Ozen(2009)}]{Gozde2009}
\bibinfo{author}{Gurdeniz, G.}, \& \bibinfo{author}{Ozen, B.}
  (\bibinfo{year}{2009}).
\newblock \bibinfo{title}{Detection of adulteration of extra-virgin olive oil
  by chemometric analysis of mid-infrared spectral data}.
\newblock {\it \bibinfo{journal}{Food Chemistry}\/},  {\it
  \bibinfo{volume}{116}\/}, \bibinfo{pages}{519--525}.
\bibitem[{Huo et~al.(2012)Huo, Jin \& Zhang}]{Quangong20122}
\bibinfo{author}{Huo, Q.}, \bibinfo{author}{Jin, X.-B.}, \&
  \bibinfo{author}{Zhang, H.} (\bibinfo{year}{2012}).
\newblock \bibinfo{title}{Multi-label classification for oil authentication}.
\newblock In {\it \bibinfo{booktitle}{9th International Conference on Fuzzy
  Systems and Knowledge Discovery}\/} (pp. \bibinfo{pages}{711--714}).
\bibitem[{Jin et~al.(2012)Jin, Geng, Yu \& Zhang}]{Xiaobo2012}
\bibinfo{author}{Jin, X.-B.}, \bibinfo{author}{Geng, G.}, \bibinfo{author}{Yu,
  J.}, \& \bibinfo{author}{Zhang, D.} (\bibinfo{year}{2012}).
\newblock \bibinfo{title}{Multi-label learning vector quantization algorithm}.
\newblock In {\it \bibinfo{booktitle}{21st International Conference on Pattern
  Recognition}\/} (pp. \bibinfo{pages}{2140--2143}).
\bibitem[{Lees(2003)}]{Lees2003}
\bibinfo{author}{Lees, M.} (\bibinfo{year}{2003}).
\newblock {\it \bibinfo{title}{Food authenticity and traceability}\/}.
\newblock \bibinfo{publisher}{Woodhead Publishing in Food Science and
  Technology}.
\bibitem[{Lerma-Garc\'{\i}a et~al.(2008)Lerma-Garc\'{\i}a, Ramis-Ramos,
  Herrero-Mart\'{\i}nez \& Sim\'{o}-Alfonso}]{Maria2008}
\bibinfo{author}{Lerma-Garc\'{\i}a, M.~J.}, \bibinfo{author}{Ramis-Ramos, G.},
  \bibinfo{author}{Herrero-Mart\'{\i}nez, J.~M.}, \&
  \bibinfo{author}{Sim\'{o}-Alfonso, E.~F.} (\bibinfo{year}{2008}).
\newblock \bibinfo{title}{Classification of vegetable oils according to their
  botanical origin using sterol profiles established by direct infusion mass
  spectrometry}.
\newblock {\it \bibinfo{journal}{Rapid Communications in Mass Spectrometry}\/},
   {\it \bibinfo{volume}{22}\/}, \bibinfo{pages}{973--978}.
\bibitem[{Lizhi et~al.(2010)Lizhi, Toyoda \& Ihara}]{Hu2010}
\bibinfo{author}{Lizhi, H.}, \bibinfo{author}{Toyoda, K.}, \&
  \bibinfo{author}{Ihara, I.} (\bibinfo{year}{2010}).
\newblock \bibinfo{title}{Discrimination of olive oil adulterated with
  vegetable oils using dielectric spectroscopy}.
\newblock {\it \bibinfo{journal}{Journal of Food Engineering}\/},  {\it
  \bibinfo{volume}{96}\/}, \bibinfo{pages}{167--171}.
\bibitem[{Lorenzo et~al.(2002)Lorenzo, Pavon, Laespada, Pinto \&
  Cordero}]{Isabel2002}
\bibinfo{author}{Lorenzo, I.~M.}, \bibinfo{author}{Pavon, J. L.~P.},
  \bibinfo{author}{Laespada, M. E.~F.}, \bibinfo{author}{Pinto, C.~G.}, \&
  \bibinfo{author}{Cordero, B.~M.} (\bibinfo{year}{2002}).
\newblock \bibinfo{title}{Detection of adulterants in olive oil by
  headspace¨Cmass spectrometry}.
\newblock {\it \bibinfo{journal}{Journal of Chromatography}\/},  {\it
  \bibinfo{volume}{945}\/}, \bibinfo{pages}{221--230}.
\bibitem[{Maggio et~al.(2010)Maggio, Cerretani, Chiavaro, Kaufman \&
  Bendini}]{Ruben2010}
\bibinfo{author}{Maggio, R.~M.}, \bibinfo{author}{Cerretani, L.},
  \bibinfo{author}{Chiavaro, E.}, \bibinfo{author}{Kaufman, T.~S.}, \&
  \bibinfo{author}{Bendini, A.} (\bibinfo{year}{2010}).
\newblock \bibinfo{title}{A novel chemometric strategy for the estimation of
  extra virgin olive oil adulteration with edible oils}.
\newblock {\it \bibinfo{journal}{Food Control}\/},  {\it
  \bibinfo{volume}{21}\/}, \bibinfo{pages}{890--895}.
\bibitem[{Mariani et~al.(2006)Mariani, Bellan, Lestini \& Aparicio}]{Carlo2006}
\bibinfo{author}{Mariani, C.}, \bibinfo{author}{Bellan, G.},
  \bibinfo{author}{Lestini, E.}, \& \bibinfo{author}{Aparicio, R.}
  (\bibinfo{year}{2006}).
\newblock \bibinfo{title}{The detection of the presence of hazelnut oil in
  olive oil by free and esterified sterols}.
\newblock {\it \bibinfo{journal}{European Food Research and Technology}\/},
  {\it \bibinfo{volume}{223}\/}, \bibinfo{pages}{655--661}.
\bibitem[{Oliveros et~al.(2002)Oliveros, Pav¨®n, Pinto, Laespada, Cordero \&
  Forina}]{Concepcion2002}
\bibinfo{author}{Oliveros, M. C.~C.}, \bibinfo{author}{Pav¨®n, J. L.~P.},
  \bibinfo{author}{Pinto, C.~G.}, \bibinfo{author}{Laespada, M. E.~F.},
  \bibinfo{author}{Cordero, B.~M.}, \& \bibinfo{author}{Forina, M.}
  (\bibinfo{year}{2002}).
\newblock \bibinfo{title}{Electronic nose based on metal oxide semiconductor
  sensors as a fast alternative for the detection of adulteration of virgin
  olive oils}.
\newblock {\it \bibinfo{journal}{Analytica Chimica Acta}\/},  {\it
  \bibinfo{volume}{459}\/}, \bibinfo{pages}{219--228}.
\bibitem[{Oussama et~al.(2012)Oussama, Elabadi, Platikanov, Kzaiber \&
  Tauler}]{Abdelkhalek2012}
\bibinfo{author}{Oussama, A.}, \bibinfo{author}{Elabadi, F.},
  \bibinfo{author}{Platikanov, S.}, \bibinfo{author}{Kzaiber, F.}, \&
  \bibinfo{author}{Tauler, R.} (\bibinfo{year}{2012}).
\newblock \bibinfo{title}{Detection of olive oil adulteration using ft-ir
  spectroscopy and pls with variable importance of projection (vip) scores}.
\newblock {\it \bibinfo{journal}{Journal of the American Oil Chemists'
  Society}\/},  {\it \bibinfo{volume}{89}\/}, \bibinfo{pages}{1807--1812}.
\bibitem[{Rohmana \& Man(2011)}]{Abdul2011}
\bibinfo{author}{Rohmana, A.}, \& \bibinfo{author}{Man, Y. B.~C.}
  (\bibinfo{year}{2011}).
\newblock \bibinfo{title}{The use of fourier transform mid infrared (ft-mir)
  spectroscopy for detection and quantification of adulteration in virgin
  coconut oil}.
\newblock {\it \bibinfo{journal}{Food Chemistry}\/},  {\it
  \bibinfo{volume}{129}\/}, \bibinfo{pages}{583--588}.
\bibitem[{Schapire \& Singer(1999)}]{Schapire1999}
\bibinfo{author}{Schapire, R.~E.}, \& \bibinfo{author}{Singer, Y.}
  (\bibinfo{year}{1999}).
\newblock \bibinfo{title}{Improved boosting algorithms using confidence-rated
  predictions}.
\newblock {\it \bibinfo{journal}{Machine Learning}\/},  {\it
  \bibinfo{volume}{37}\/}, \bibinfo{pages}{297--336}.
\bibitem[{Sikorska et~al.(2005)Sikorska, G\'{o}recki, Khmelinskii, Sikorski \&
  Kozio{\l}}]{Ewa2005}
\bibinfo{author}{Sikorska, E.}, \bibinfo{author}{G\'{o}recki, T.},
  \bibinfo{author}{Khmelinskii, I.~V.}, \bibinfo{author}{Sikorski, M.}, \&
  \bibinfo{author}{Kozio{\l}, J.} (\bibinfo{year}{2005}).
\newblock \bibinfo{title}{Classification of edible oils using synchronous
  scanning fluorescence spectroscopy}.
\newblock {\it \bibinfo{journal}{Food Chemistry}\/},  {\it
  \bibinfo{volume}{89}\/}, \bibinfo{pages}{217--225}.
\bibitem[{Tsoumakas et~al.(2010)Tsoumakas, Katakis \& Vlahavas}]{Tsoumakas2010}
\bibinfo{author}{Tsoumakas, G.}, \bibinfo{author}{Katakis, I.}, \&
  \bibinfo{author}{Vlahavas, I.} (\bibinfo{year}{2010}).
\newblock \bibinfo{title}{Mining multi-label data}.
\newblock {\it \bibinfo{journal}{Data Mining and Knowledge Discovery
  Handbook}\/},  (pp. \bibinfo{pages}{667--685}).
\bibitem[{Vigli et~al.(2003)Vigli, Philippidis, Spyros \& Dais}]{Vigli2003}
\bibinfo{author}{Vigli, G.}, \bibinfo{author}{Philippidis, A.},
  \bibinfo{author}{Spyros, A.}, \& \bibinfo{author}{Dais, P.}
  (\bibinfo{year}{2003}).
\newblock \bibinfo{title}{Classification of edible oils by employing 31p and 1h
  nmr spectroscopy in combination with multivariate statistical analysis. a
  proposal for the detection of seed oil adulterated in virgin olive oils}.
\newblock {\it \bibinfo{journal}{Journal of agricultural and food
  chemistry}\/},  {\it \bibinfo{volume}{51}\/}, \bibinfo{pages}{5715--5722}.
\bibitem[{Wu et~al.(2008)Wu, He, Feng \& Sun}]{Wu2008}
\bibinfo{author}{Wu, D.}, \bibinfo{author}{He, Y.}, \bibinfo{author}{Feng, S.},
  \& \bibinfo{author}{Sun, D.-W.} (\bibinfo{year}{2008}).
\newblock \bibinfo{title}{Study on infrared spectroscopy technique for fast
  measurement of protein content in milk powder based on ls-svm}.
\newblock {\it \bibinfo{journal}{Journal of Food Engineering}\/},  {\it
  \bibinfo{volume}{84}\/}, \bibinfo{pages}{124--131}.
\bibitem[{Yang et~al.(2005)Yang, Irudayara \& Paradkar}]{Hong2005}
\bibinfo{author}{Yang, H.}, \bibinfo{author}{Irudayara, J.}, \&
  \bibinfo{author}{Paradkar, M.~M.} (\bibinfo{year}{2005}).
\newblock \bibinfo{title}{Discriminant analysis of edible oils and fats by
  ftir, ft-nir and ft-raman spectroscopy}.
\newblock {\it \bibinfo{journal}{Food Chemistry}\/},  {\it
  \bibinfo{volume}{93}\/}, \bibinfo{pages}{25--32}.

\end{thebibliography}

\end{document}